\documentclass[useAMS,usenatbib,usegraphicx,referee]{mnras}
\usepackage{graphicx,amssymb,amsmath,times,verbatim}
\def\lta{\mathrel{\spose{\lower 3pt\hbox{$\mathchar"218$}}
     \raise 2.0pt\hbox{$\mathchar"13C$}}}
\def\gta{\mathrel{\spose{\lower 3pt\hbox{$\mathchar"218$}}
     \raise 2.0pt\hbox{$\mathchar"13E$}}}

\def\mathnew{\mathsurround=0pt}

\def\simov#1#2{\lower .5pt\vbox{\baselineskip0pt \lineskip-.5pt
\ialign{$\mathnew#1\hfil##\hfil$\crcr#2\crcr\sim\crcr}}}

\title[Multi-wavelength Variability of OJ 287]{Multi-wavelength temporal and spectral variability of the
blazar OJ 287 during and after the December 2015 flare: a major accretion disc contribution}

\author[Kushwaha et al.]{Pankaj Kushwaha$^{1}$\thanks{E-mail: pankaj.kushwaha@iag.usp.br},
Alok C. Gupta$^{2,3}$\thanks{E-mail: acgupta30@gmail.com}\thanks{CAS PIFI Fellow}, 
Paul J. Wiita$^{4}$\thanks{Email: wiitap@tcnj.edu}, Haritma Gaur$^{2}$,  
\newauthor 
E.~M. de Gouveia Dal Pino$^{1}$, Jai Bhagwan$^{5}$, O.~M.~Kurtanidze$^{6,7}$, V.~M.~Larionov$^{8,9}$,
\newauthor 
G.~Damljanovic$^{10}$, M.~Uemura$^{11}$, E.~Semkov$^{12}$, A.~Strigachev$^{12}$, R.~Bachev$^{12}$,  
O.~Vince$^{10}$,
\newauthor 
Minfeng Gu$^{2}$, Z. Zhang$^{13}$,
T. Abe$^{11}$, A. Agarwal$^{3}$, G.~A.~Borman$^{14}$, J. H. Fan$^{15}$, 
\newauthor 
T.~S.~Grishina$^{8}$, J. Hirochi$^{11}$, R. Itoh$^{16}$, M. Kawabata$^{11}$, E.~N.~Kopatskaya$^{8}$, 
\newauthor 
S. O. Kurtanidze$^{6}$, E.~G.~Larionova$^{8}$, L.~V.~Larionova$^{8}$, A. Mishra$^{3}$, D.~A.~Morozova$^{8}$, 
\newauthor 
T. Nakaoka$^{11}$, M. G. Nikolashvili$^{6}$, S.~S.~Savchenko$^{8}$, Yu.~V.~Troitskaya$^{8}$, 
\newauthor 
I.~S.~Troitsky$^{8}$, A.~A.~Vasilyev$^{8}$  \\
\\
$^{1}$ Department of Astronomy (IAG-USP), University of Sao Paulo, Sao Paulo 05508-090, Brazil \\
$^{2}$ Key Laboratory for Research in Galaxies and Cosmology, Shanghai Astronomical
Observatory, Chinese Academy of Sciences, \\
~~~80 Nandan Road, Shanghai 200030, China \\
$^{3}$ Aryabhatta Research Institute of Observational Sciences (ARIES), Manora Peak, Nainital 263002, India \\
$^{4}$ Department of Physics, The College of New Jersey, P.O.\ Box 7718, Ewing, NJ 08628-0718, USA  \\
$^{5}$ School of Studies in Physics \& Astrophysics, Pt Ravishankar Shukla University, Amanaka G.E.\ Road, 
Raipur 492010, India \\
$^{6}$Abastumani Observatory, Mt. Kanobili, 0301 Abastumani, Georgia \\
$^{7}$Engelhardt Astronomical Observatory, Kazan Federal University, Tatarstan, Russia \\
$^{8}$Astronomical Institute, St.-Petersberg State University, 198504 St.-Petersberg, Russia \\
$^{9}$Pulkovo Observatory, 196140 St.-Petersberg, Russia \\
$^{10}$Astronomical Observatory, Volgina 7, 11060 Belgrade, Serbia \\
$^{11}$Hiroshima Astrophysical Science Center, Hiroshima University, Kagamiyama 1-3-1, Higashi-Hiroshima 739-8526, Japan \\
$^{12}$Institute of Astronomy and National Astronomical Observatory, Bulgarian Academy of Sciences, 72 Tsarigradsko 
Shosse Blvd., 1784 Sofia, Bulgaria \\
$^{13}$Shanghai Astronomical Observatory, Chinese Academy of Sciences, 80 Nandan Road, Shanghai 200030, China \\
$^{14}$Crimean Astrophysical Observatory, P/O Nauchny, Crimea, 298409, Russia \\
$^{15}$Center for Astrophysics, Guangzhou University, Guangzhou 510006, China \\
$^{16}$Department of Physics, Tokyo Institute of Technology, 2-12-1 Ookayama, Meguro-ku, Tokyo 152-8551, Japan \\
}

\begin{document}

\maketitle

\begin{abstract}
We present a multi-wavelength spectral and temporal analysis of the blazar OJ 287
during its recent activity between December 2015 -- May 2016, showing strong
variability in the near-infrared (NIR) to X-ray energies with detection at
$\gamma$-ray energies as well. Most of the optical flux variations exhibit strong changes in polarization
angle and degree. All the inter-band time lags are consistent with
simultaneous emissions. Interestingly, on days with excellent data coverage
in the NIR--UV  bands, the  spectral energy distributions (SEDs)  show signatures of bumps in
the visible--UV bands, never seen before in this source. The optical bump can
be explained as accretion-disk emission associated with the primary black hole of mass $\sim \rm 1.8
\times10^{10}  M_{\odot}$ while the little bump feature in the optical-UV appears
consistent with line emission. Further, the broadband SEDs extracted during the
first flare and during a quiescent period during this span show very different
$\gamma$-ray spectra compared to previously observed flare or quiescent spectra.
The probable thermal bump in the visible seems to have been clearly present since
May 2013, as found by examining all available NIR-optical observations, and favors
the binary super-massive black hole model. The simultaneous multi-wavelength
variability and relatively weak $\gamma$-ray emission that shows a shift
in the SED peak is consistent with $\gamma$-ray emission originating
from inverse Compton scattering of photons from the line emission that
apparently contributes to the little blue bump.

\end{abstract}

\begin{keywords}
 BL Lac objects: individual: OJ 287 -- galaxies: active --
-- galaxies: jets -- radiation mechanisms: non-thermal -- X-rays: galaxies
\end{keywords}

\section{Introduction} \label{sec:intro}
OJ 287 belongs to the BL Lacertae subclass of blazars, a class of radio-loud active
galactic nuclei (AGN) with relativistic jets aligned at small angles to the observer's
line of sight and with very weak or no emission lines in their optical-ultraviolet
spectra. The emission is almost entirely non-thermal, spanning the full observable
electromagnetic spectrum, and is characterized by strong and rapid variability with
high radio-to-optical polarization and frequent detections of superluminal features
\citep{2013AJ....146..120L}. Blazar variability is observed in all domains:
temporal, spatial, spectral and polarimetric, and covers a wide range in each. In the
time domain, the variability across the spectrum is erratic and stochastic, and is
seen on all timescales from minutes to decades. The spectral domain, on the other
hand, shows the characteristic broad double humped spectral energy distributions
(SEDs) with a low energy hump peaking between infra-red (IR) and ultraviolet, 
or sometimes, soft X-ray bands, while the high energy hump reaches its maximum
at $\gamma$-ray energies \citep{1998MNRAS.299..433F,2016ApJS..224...26M}. The emission
at the low energy hump is understood to be the synchrotron emission from
relativistic non-thermal electrons in the jet owing to the high degree of polarization
in radio and optical bands, which often changes with the flux state of the
source. The origin of the high energy emission, on the other hand, is less certain;
while it probably results from  inverse Compton (IC) processes, a hadronic origin is
also plausible \citep[e.g.][]{2001APh....15..121M}.

OJ 287 ($z = 0.306$) is one of the brightest and most highly variable BL Lac objects
at radio and optical energies. It is also one of the most extensively observed
extragalactic objects, with optical R-band data available since 1890. Despite blazars
being known for stochastic temporal variations, an extraordinary property of this
source is the occurrence of  $\sim 12$ year nearly periodic outbursts in its optical flux
\citep{1996A&A...305L..17S}. Ever since this finding and its likely explanation in
terms of a binary super-massive black hole (SMBH) system where the major flares are
attributed to the secondary SMBH striking the accretion disc around the primary one
\citep[but see \citet{2010MNRAS.402.2087V}]{1988ApJ...325..628S}, it has been
a  prime target for multi-wavelength observations.  OJ 287 has been monitored both
intensively and extensively across the electromagnetic spectrum during the times
of its expected outbursts  \citep[and references therein]{2011AcPol..51f..76V,
2016ApJ...819L..37V, 2017MNRAS.465.4423G,2017ApJ...835..275R}. The early intensive
observations revealed a double flare \citep[see also \citet{2013A&A...559A..20H}]
{1996A&A...315L..13S} and the SMBH binary model was refined and improved by incorporating
inputs from these observations \citep{1996ApJ...460..207L,2006ApJ...646...36V}.
These have led
to attempts  to predict and explain the emission properties \citep[e.g.][]
{2013ApJ...764....5P,2016MNRAS.457.1145P} and derive constraints on the system
related to general relativistic effects such as loss of orbital energy through
gravitational radiation \citep{2016ApJ...819L..37V}. Apart from the $\sim$12
y periodic feature, other periodic features have been claimed  \citep{2006AJ....132.1256W,
2016ApJ...832...47B,2016AJ....151...54S,2013MNRAS.434.3122P}.  Although other models
have been proposed, none of them have been as successful as the binary SMBH picture
in term of predicting the occurrence of the flares at $\sim$ 12 years periodicity 
\citep[and references therein]{2012MNRAS.427...77V,2011AcPol..51f..76V};
also see \cite{2010MNRAS.402.2087V}.

Apart from these photometric observations, the jet kinematics, multi-wavelength
light curves, and polarization properties of OJ 287 have been studied in detail
to understand the jet properties and the locations of flaring regions. Studies
at radio energies using high resolution Very Large Baseline Array (VLBA) radio
maps suggest a wobbling parsec-scale jet \citep{2004ApJ...608..149T,2011ApJ...726L..13A,
2011AJ....141..178M} with changes in projected jet-position-angle of $> 100^\circ$
\citep{2017A&A...597A..80H}. In the optical, OJ 287 shows a preferred position angle
in polarization, indicating the presence  of both a steady core and a chaotic jet
component \citep{2010MNRAS.402.2087V}. On the other hand, correlation studies between
$\gamma$-ray and radio emission show significant correlations with flaring episodes
often associated with the ejection of superluminal components from stationary knots,
located at parsec scales down the jet and seen in VLBA maps \citep{2011ApJ...726L..13A,
2015PKAS...30..429S,2017A&A...597A..80H}. A similar location of a flaring region was
inferred from the broadband SED modeling during a 2009 flare by \cite{2013MNRAS.433.2380K}.

Here, we present a correlation analysis of multi-wavelength (MW) observations of
OJ 287 made before, during, and after its recent flaring in December 2015, which was predicted
by the binary SMBH model \citep{2011ApJ...742...22V}, and claimed to be the
result of the impact of the secondary SMBH on the accretion disk of the primary.
\citet{2016ApJ...819L..37V} has already shown that the timing of this flare can
provide a remarkably improved value of primary SMBH spin to $\rm 0.313 \pm
0.01$, apart from producing values for the mass of the primary SMBH to be $\rm
(1.83 \pm 0.01) \times 10^{10} M_{\odot}$ and that of the secondary to be $\rm (1.5
\pm0.1) \times 10^{8} M_{\odot}$, along with an orbital eccentricity of $0.700\pm0.001$.  
Additionally, they also performed a preliminary modeling of the light curves and
degree of polarization following \citet{2016MNRAS.457.1145P} and suggested that the
X-ray emission is primarily non-thermal in origin. Intensive observations and intra-night
variability during this period at optical energies have also been presented along
with polarization data \citep{2017MNRAS.465.4423G,2017ApJ...835..275R}. Our focus
here is on what may be learned from MW emission and trying to understand the putative
impact's effect on MW emissions.  We note that our observations extend through 2016 May, well beyond
those discussed in \citet{2016ApJ...819L..37V}.

The paper is organized into five sections with \S2 summarizing the MW data we
have compiled and the procedures associated with them. The MW cross-correlation
analyses and  SEDs details and results are presented in \S3. The implications
of these results and a discussion is presented in \S4, with our conclusions
given in \S5. A $\Lambda$CDM cosmology with $\rm H_0$ = 69.6 km s$^{-1}$
Mpc$^{-1}$, $\rm \Omega_M$ = 0.286 and $\Omega_\Lambda$ = 0.714 is assumed for
calculation of physical quantities.

\section{Multi-wavelength Data and Reduction} \label{sec:data}
The MW data used here belongs both to publicly available data archives (X-ray,
$\gamma$-ray. radio and a part of NIR-optical-UV measurements) as well as observations we made
in the optical/NIR at specific observatories across the globe.

\subsection{Fermi $\gamma$-ray Data}

The $\gamma$-ray data belong to the \emph{Fermi}-LAT (Large Area Telescope),
an imaging telescope sensitive to photons with energies from 20 MeV to $>$ 300
GeV \citep{2009ApJ...697.1071A}. Except for TOO observations, it is normally
operated in all-sky scanning mode, covering the entire sky every $\sim$3 hours.
The LAT data being used here belong to one such normal operation mode from 
15 October 2015 to 24 May  2016 (MJD: 57310 -- 57560). The data were analyzed following
the recommended procedures for the latest \emph{PASS8} instrument response function
data with the \emph{Fermi Science Tool} version v10r0p5.

Our analysis considered only events between energies 0.1$\leq$E$\leq$300 GeV, categorized
as ``evclass=128, evtype=3'' and within the maximum zenith angle of 90$^\circ$ from
a circular region of interest (ROI) of $15^\circ$ centered on the source. These
events were further filtered with the appropriate good time intervals using the
recommended criteria ``(DATA\_QUAL$>$0)\&\&(LAT\_CONFIG==1)'' to ensure that the
instrument operation was in the normal scientific mode. The likely effects
of these cuts and selections, as well as the presence of other sources in the ROI,
were incorporated by generating an exposure map on the ROI and an additional annulus
of $10^\circ$ around it. We then modeled these events  using `unbinned likelihood
analysis' with an input source model file from the 3rd LAT catalog
\citep[3FGL --gll\_psc\_v16.fit;][]{2015ApJS..218...23A}. The Galactic and isotropic
extragalactic contributions were accounted for by using the respective templates,
\emph{gll\_iem\_v06.fits} and \emph{iso\_P8R2\_SOURCE\_V6\_v06.txt} file provided
by the instrument team. A significance criterion of $3\sigma$, corresponding to a
TS (Test Statistic) value of $\sim10$ has been used for source detection.

We generated the daily $\gamma$-ray light curve following the above-mentioned
procedures. The source was modeled with a log-parabola $\gamma$-ray spectrum while
the default models, as used in the 3FGL catalog, were used for the rest of the sources
in the input model file. For SEDs, the selected time intervals were further divided into
logarithmic energy bins and a power-law model was used for each bin to extract the
source flux. The $\gamma$-ray light curve is given in the top panel of Figure \ref{fig:mwlc}.

\subsection{Swift X-ray and UV/Optical Data}

{\bf X-ray:} All the \emph{Swift}-XRT observations taken during this period for
OJ 287  were made in the photon counting (PC) pointed mode. Accordingly, the data
files were first processed using the latest \emph{CALDB} files via the \emph{xrtpipeline}
task with the default parameters. The events associated with the source and
background regions, suitable for estimation of physical quantities of interest, were
extracted using the \emph{xselect} task with a circular source region of $\rm 47.2''$
\citep[90\% PSF,][]{2005SPIE.5898..360M} and an annular region free of source contamination,
respectively. None of the observations were found to have a count rate $\gtrsim$ 0.5
counts/s, which would have required pile-up corrections. The effects of selections,
cuts, and instrument operation related biases were estimated by generating an ancillary
response file through the \emph{xrtmkarf} task. Finally, we derived the flux assuming
a power-law model with neutral hydrogen (nH) column density along the source direction
fixed to the Galactic value of $\rm 2.38 \times 10^{20}~ cm^{-2}$ using
the corresponding axillary files within \emph{XSPEC}. The second panel of Figure
\ref{fig:mwlc} displays the \emph{Swift} X-ray data reduced in this fashion.

\vspace*{0.05in}
\noindent
{\bf UV/Optical:}   
We have analyzed all the observations of the source OJ 287 performed by the {\it Swift}
Ultraviolet and Optical Telescope \citep[UVOT;][]{2005SSRv..120...95R} between MJD 57354
and 557535. In this study we have utilized only level 2 image mode data from UVOT
in all filters (V, B, U, UVW1, UVM2, UVW2) in which the image is directly accumulated
onboard, discarding the photon timing information so as to reduce the telemetry
volume. We have used the standard procedure given on the {\it Swift}
website\footnote{http://www.swift.ac.uk/analysis/uvot/} using the {\it uvotmaghist}
task of the {\it heasoft} package for data reduction. {\it Swift}/UVOT source counts
were extracted from a circular region of radius 5 arcsec, centred on the source
position. The background counts were estimated from a nearby larger, source free,
circular region of radius 20 arcsec. The source OJ 287 fluxes corresponding to the
six optical/UV filters were corrected for Galactic reddening \citep[E(B$-$V) = 0.019
mag;][]{1998ApJ...500..525S}. We have used equation (2) of \citet{2009ApJ...690..163R}
to calculate  the galactic extinction corrected flux of each spectral band. The third
panel of Figure \ref{fig:mwlc} shows the {\it Swift}/UVOT light curves.

\subsection{Optical and Near-Infrared Data}

{\bf Photometry:} The bulk of the optical and near-infrared (NIR) photometric data
we display here were taken at 11 observatories around the world in B, V, R, I, J and
K bands, though the coverage is best in the R, B and V bands. The details of the
telescopes from which we collected data in  Bulgaria (3 telescopes), Chile
(SMARTS\footnote{http://www.astro.yale.edu/smarts/glast/tables/OJ287.tab}), Georgia,
India, Japan, Serbia,  and USA (2 telescopes), and how those data were reduced  are
presented in \citet{2017MNRAS.465.4423G}.  We have now included photometric data
from St.\ Petersburg University from the 70 cm telescope in Crimea and the 40 cm
telescope in St.\ Petersburg Observatory, both equipped with nearly  identical
imaging photometers--polarimeters. The reduction of those observations is discussed
in \citet{2008A&A...492..389L}. 

\begin{figure}
\begin{center}
\includegraphics[scale=1]{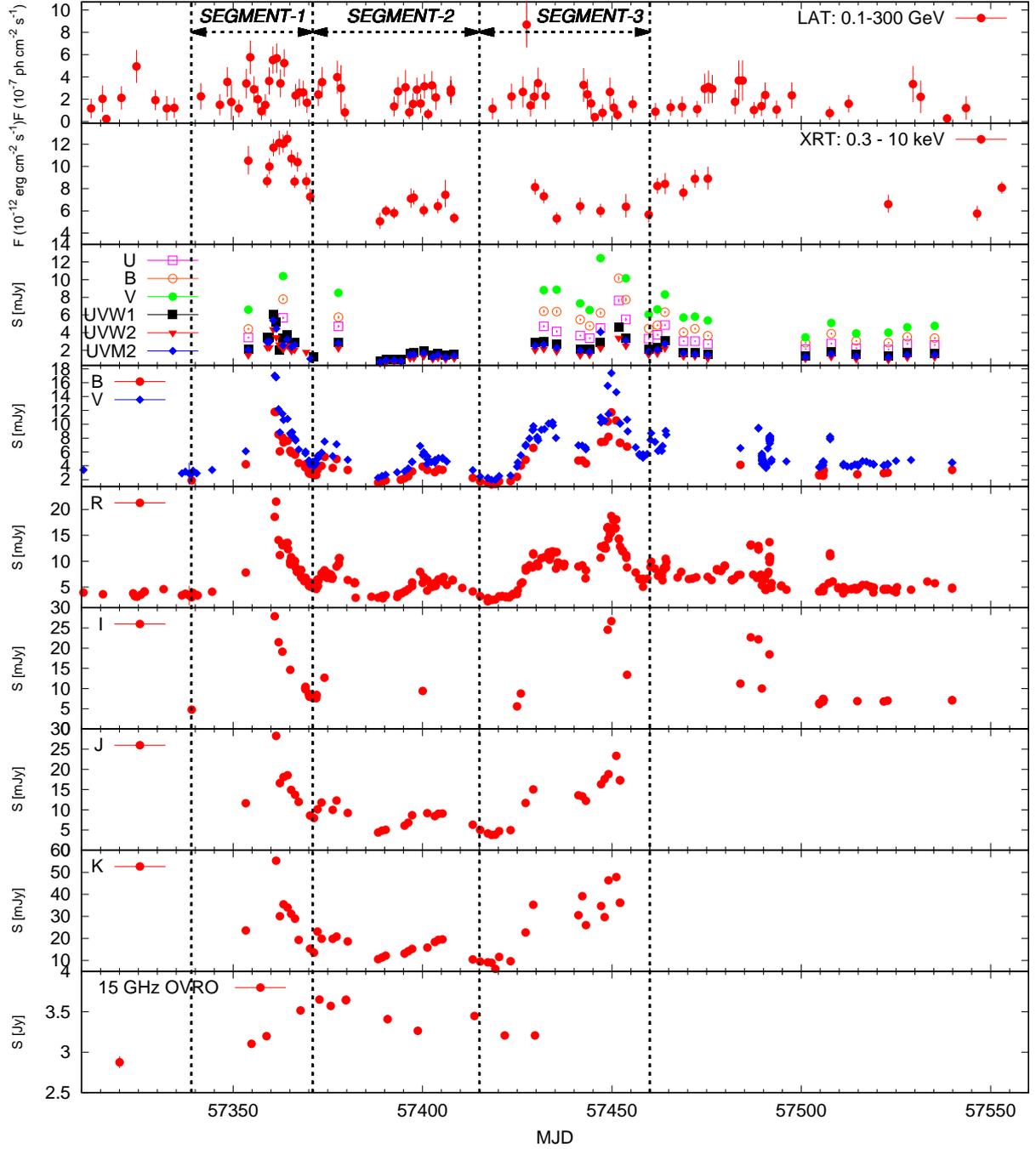}
\end{center}
\caption{October 2015 -- May 2016 multi-wavelength light curves of OJ 287 from $\gamma$-rays
to radio. The vertical lines demarcate the sections considered for cross-correlation analyses
in the present work.}
\label{fig:mwlc}
\end{figure}

The photometric data were corrected for extinction following \citet{2011ApJ...737..103S}
with an E(B-V) value of $\rm 0.0280 \pm 0.0008$ while the magnitude to flux conversion
was performed using the zero point fluxes from \citet{1998A&A...333..231B}. These
data are shown in the fourth through eighth rows of Figure \ref{fig:mwlc}. Although
we do not display separate symbols for different observatories, one can see from
\cite{2017MNRAS.465.4423G} that whenever there are multiple measurements made at
close times the agreements are excellent, and the same holds true for the new optical
data that has been incorporated in this paper.

{\bf Polarimetry:} We have obtained a substantial amount of  polarimetric data
from the St.\  Petersberg and Crimean telescopes.  Polarimetric observations were
performed using two Savart plates rotated by 45{\degr} relative to each another.
By swapping the plates, the observer can obtain the relative Stokes $q$ and $u$
parameters from the two split images of each source in the field \citep{2008A&A...492..389L}.
Instrumental polarization was found via stars located near the object under the
assumption that their radiation is unpolarized. This is indicated also by the low
level of extinction in the direction of OJ~287 \citep[$A_V=0.077$ mag;][]{2011ApJ...737..103S}.
        
Additional polarimetric data for the same period have been taken from the  public
archive of the Steward Observatory of the University of Arizona. The details of this
program and analysis procedures are discussed in \citet{2009arXiv0912.3621S}\footnote{http://james.as.arizona.edu/$\sim$psmith/Fermi}.
 These polarization data are presented in the second and third panels of Figure \ref{fig:pollc}.
We note that during periods of overlap the fractional polarization measurements
agree extremely well, as do the polarization angles when the $\pm n \pi$ ambiguity
is taken into consideration.

\subsection{Radio data}  

Radio data at 15 GHz are taken from the public archive of Owens Valley Radio Observatory
(OVRO). The observations were made using the 40 meter single dish radio telescope
under the project {\it Monitoring of Fermi Blazars}\footnote{http://www.astro.caltech.edu/ovroblazars/}.
The OVRO telescope monitors Fermi detected blazars north of 20$^\circ$ declination
\citep{2008ApJS..175...97H}. The bottom panel of Figure \ref{fig:mwlc} displays
these radio data.

\subsection{Multi-wavelength Light Curves}\label{sec:analysis}

The resulting MW light curves are shown in Figure \ref{fig:mwlc}. Significant variability
can be seen all across the electromagnetic spectrum with the optical-IR showing
very prominent variations. The X-ray band is not very well sampled, but within the
span of observations, it appears to reflect the changes seen in the optical-IR light
curves, at least during the first half of the period. The $\gamma$-ray band also
shows significant variations, but to the eye, they appear to be only roughly correlated
with the optical-IR bands, while the  sparsely sampled radio data show relatively smooth
variations without any obvious connection to the other bands.

\begin{figure}
\begin{center}
\includegraphics[scale=0.8]{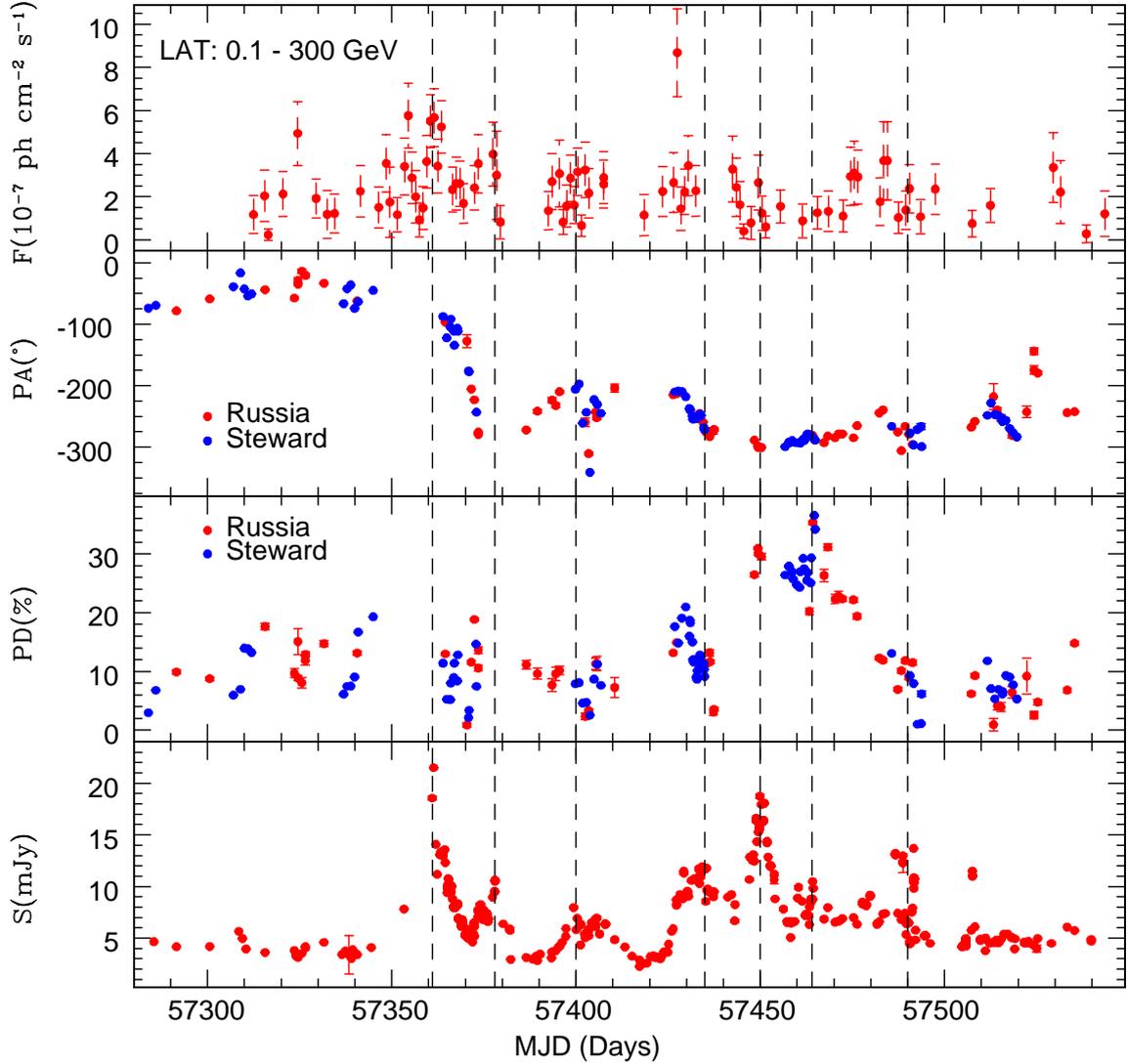}
\end{center}
\caption{September  2015 -- May 2016 optical polarization data plotted along with
the $\gamma$-ray (top) and R-band (bottom) light curves of OJ 287.  The dashed vertical lines
note the peaks of the flares discussed in the text.}
\label{fig:pollc}
\end{figure}

The polarization data show that there are at least two distinctly different types
of flares coming from  the source. The main flare at MJD 57361 which, according to the
SMBH binary model, corresponds to the impact of the secondary on the primary's disk,
evinces a relatively low polarization fraction (PD $\lesssim 11$\%), that  differs
little from that in the preceding quiescent period \citep[see Fig. 3 in][]{2016ApJ...819L..37V}.
During its decay phase, the polarization angle (PA) undergoes a large swing
of $\gtrsim 200^\circ$ while the PD varies quite a bit (Fig.\ \ref{fig:pollc}),
but other new flares appear soon thereafter. Almost all these flares until the
second strong optical-NIR flare on MJD 57450 are associated with significant change in
PA ($\gtrsim 90^\circ$) and PD both.  We lack sufficient
polarization coverage during the significant flare peaking at MJD 57378 but it is known to be
highly polarized \citep[30--40\%; Fig. 3 of ][]{2016ApJ...819L..37V}. On the other hand,
the flare around MJD 57400 is similar in its polarization properties to the strongest flare
around MJD 57361, with a significant swing in PA, but only small fluctuations around a modest
value of PD. The subsequent flares at MJD 57435  and the very large one at MJD 57450
both show strong  increments in PD, but the former corresponds to a slower swing in
PA  while the latter shows nothing detectable in that regard. We note that the flare at
MJD 57435 does correspond to a single point excursion in the $\gamma$-ray flux.
However, the strongest values of PD, exceeding 30\%, are measured around MJD 57464
when the PA is nearly constant and only a small flare in overall flux is seen. The
final significant flare in this period is at  MJD 57490 and again  the PD is
low and variable, but the PA is essentially steady. It is certainly reasonable to
take the flares with low PD to arise from predominantly thermal processes, while
those with high PD may be associated with non-thermal, presumably jet-based,
 fluctuations \citep[see also][]{2016ApJ...819L..37V}.

\section{Spectral and Temporal Variability Analysis and Results}

To better understand the MW emission of OJ 287, we have performed cross-correlations
between different bands and have examined the SEDs as a function of time. The former
analysis contains imprints of particle acceleration and radiative
cooling in different bands, while the latter helps us ascertain the nature of the
emission processes and perhaps separate thermal from non-thermal emission components.

\subsection{MW Cross-correlation Analysis} \label{subsec:zdcf}
At the start of our observing campaign in September 2015, OJ 287 was in a low
flux state. A strong outburst event started on November 14, 2015 which peaked on
December 5, 2015 (MJD 57361). It is the first predicted outburst of the binary SMBH model
\citet{2011ApJ...742...22V} and was detected and discussed by \citet{2016ApJ...819L..37V}.
After this outburst until the end of our monitoring campaign, which lasted through
May 2016, the blazar showed multiple flares at  intervals averaging $\sim$25 days.
So the source  is seen in several flare, pre-flare, and post-flare states
over the course of eight months.

To look for lags between MW bands we performed a cross-correlation analysis  using the 
\emph{z-transformed discrete correlation function} (ZDCF) method \citep{1997ASSL..218..163A,
2013arXiv1302.1508A} which works with both regularly and irregularly sampled data. 
In case of irregular sampling, the effects of non-uniformity and sparse sampling are
accounted for by using equal population binning and Fisher's $z$-transform. Since by
default, the method uses 11 data pairs per time bin, the duration for the correlation
analysis chosen in this work is a compromise between the available number of data points and
the different variability characteristics of OJ 287 across bands. To estimate the
cross-correlation coefficients, the time lag bins are constructed by making
all possible pairs of data points from one light curve with the data points of the
other.  These are then ordered based on the pairs' time differences. This ordered
set is then binned, taking a minimum of 11 pairs per bin while discarding interdependent
pairs from each bin. The newly formed bins correspond to a lag of the mean of the
binned pairs \citep[for more details see][]{2013arXiv1302.1508A}. The coefficients
are then calculated from these bins. For estimations of errors on the correlation
coefficients, a Monte Carlo simulation of the light curve is performed using information
about the observational errors on the fluxes. For each pair of simulated light curves,
 the correlation coefficients are estimated and then transformed to the $z$-space
via \citep{1997ASSL..218..163A,2013arXiv1302.1508A}
\begin{equation}
 z = \frac{1}{2} ln\left(\frac{1+r}{1-r}\right), \qquad 
 \zeta = \frac{1}{2} ln\left(\frac{1+\rho}{1-\rho}\right), \qquad
 r = tanh(z) \nonumber
 \end{equation}
where $r$ and $\rm \rho$ are, respectively, the bin correlation coefficient and the
unknown population correlation coefficient. For $\rm \rho$, it uses an ansatz ${\rm
\rho} = r$ to estimate the mean and variance of $z$ \citep[for more details]
{2013arXiv1302.1508A}. In the $z$-space, the transformed quantities are distributed
normally. Thus, the error is estimated and finally transformed back to the correlation
space providing $1\sigma$ errors on the coefficients.  

\begin{figure}
\centering
 \includegraphics[scale=0.65]{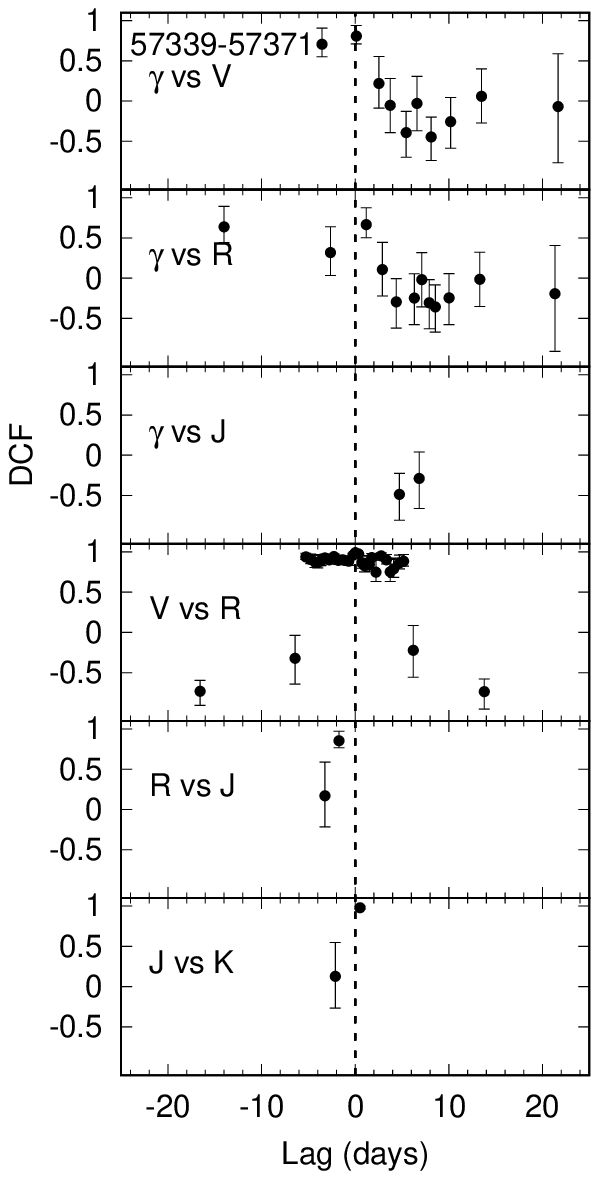}
 \includegraphics[scale=0.65]{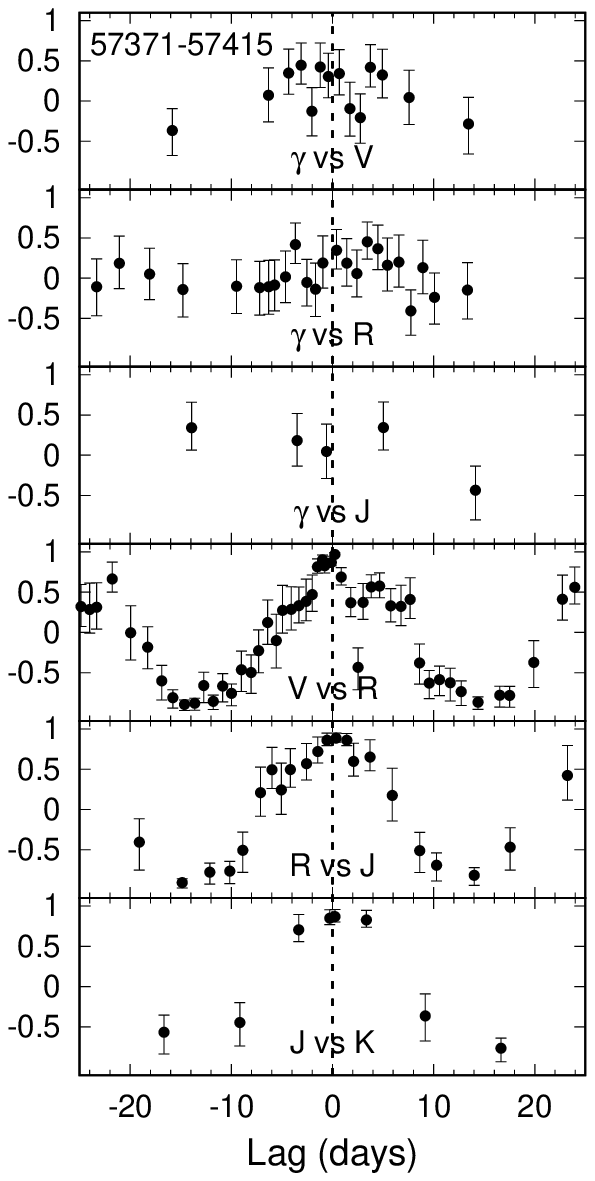}
 \includegraphics[scale=0.65]{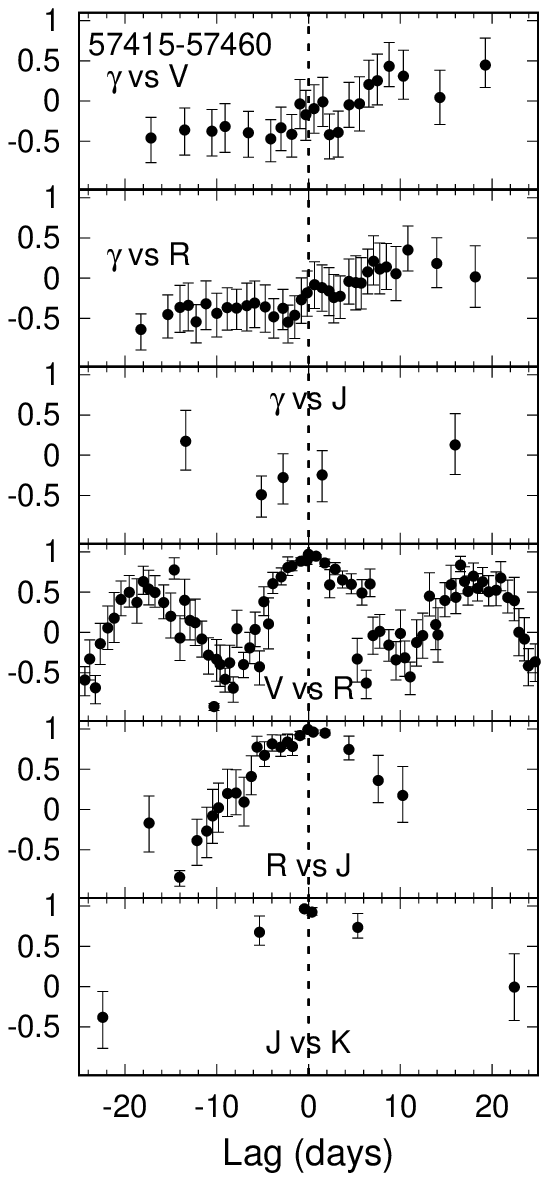}
 \includegraphics[scale=0.65]{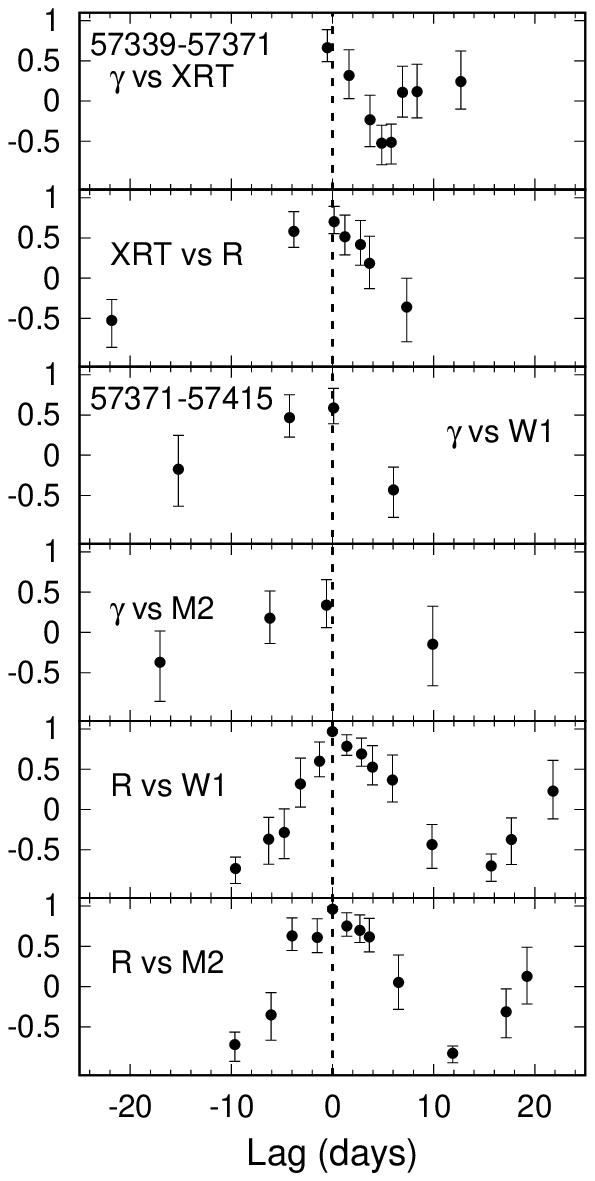}
 \caption{ DCF of $\gamma$-ray vs optical (both V and R bands); $\gamma$-ray
 vs NIR (J); optical (V) vs optical (R); NIR (J) vs NIR (K); and optical (R) vs
 NIR (J) emission for the segments  1--3 defined by the dates in each panel
 of the near continuous multi-wavelength light curves of OJ 287 (see Figure
 \ref{fig:mwlc}). The vertical dashed lines correspond to zero lag between
 the light curves labeled as ``LC1 vs LC2''. The last panel at right shows
 the DCF between other bands during these durations when data was informative
 (see \S\ref{subsec:zdcf}).}
 \label{fig:zdcf}
\end{figure}

The first three vertical sets of panels in Figure \ref{fig:zdcf} show these correlations 
plotted as DCFs of: $\gamma-$ray vs optical (both V and R bands); $\gamma-$ray vs
NIR (J); optical (V) vs optical (R); NIR (J) vs NIR (K); and optical (R) vs NIR (J). 
They were computed using the ZDCF method for three temporal segments chosen to
illustrate different phases of the light curves: MJDs 57339--57371, 57371--57415,
and 57415--57460, respectively denoted as  Segments 1, 2, and 3 in Fig. \ref{fig:mwlc}. The fourth
column of panels shows the DCFs between other bands for which we have enough measurements
during some of these segments. The errors on the coefficients are derived from 1000
realizations of each light curve pair. The corresponding lag values are reported in
Table \ref{tab:lagResutls} where a `---' entry means not enough DCF points were available
to derive lag values. A positive lag value between light curves in two different bands, labeled ``\emph{LC1}
vs \emph{LC2}'' in  Fig. \ref{fig:zdcf} and Table \ref{tab:lagResults} means that
\emph{LC2} emission lags \emph{LC1}, while a negative value would have \emph{LC2}
leading \emph{LC1}. We note that all nominal lags are within $1 \sigma$ of 0.

\begin{table}
  \centering
  \caption{Lag results for all the segments (in days)}
  \label{tab:lagResults}
  \begin{tabular}{ccccc}
  \hline
Light curves & \emph{Segment-1$^1$} & \emph{Segment-2$^2$}   & \emph{Segment-3} & \emph{Last Column, Fig.\ 3} \\ \hline
$\gamma$ vs V & $+0.1^{+1.1}_{-2.2}$ & $-1.2^{+1.7}_{-0.4}$  & -- & $-0.5^{+1.4}_{-6.0}~\rm (\gamma~ vs~ XRT)^1$ \\ 
$\gamma$ vs R & $+1.2^{+0.9}_{-2.0}$ & $+0.4^{+1.0}_{-1.2}$ & -- & $+0.16^{+1.6}_{-2.4}~\rm (XRT ~vs ~R)^1$ \\
$\gamma$ vs J & -- 		 & $+5.0^{+4.7}_{-24.4}$ & -- & $+0.1^{+2.7}_{-2.7}~\rm (\gamma~ vs~ U1)^2$ \\
V vs R 	& $+0.02^{+0.21}_{-0.12}$ & $+0.23^{+0.28}_{-0.26}$ & $0.0^{+0.9}_{-0.1}$    & $+0.6^{+5.7}_{-3.4}~\rm (\gamma~ vs~ M2)^2$ \\
R vs J 	& -- 			& $+0.4^{+1.0}_{-1.0}$ & $-0.06^{+0.27}_{-0.38}$ & $-0.02^{+0.7}_{-0.6}~\rm (R~ vs~ U1)^2$ \\
J vs K & -- 			& $+0.2^{+3.7}_{-2.4}$ & $+0.4^{+2.1}_{-2.1}$    & $+0.01^{+0.67}_{-0.68}~\rm (R~ vs~ M2)^2$ \\
  \hline
  $^1$ for MJD 57339--57371 & $^2$ for MJD 57371--57425 & & & \\
  \end{tabular}
\label{tab:lagResutls}
 \end{table}

\subsection{Spectral Energy Distributions} \label{subsec:SED}
\begin{figure}
\centering
 \includegraphics[scale=0.83]{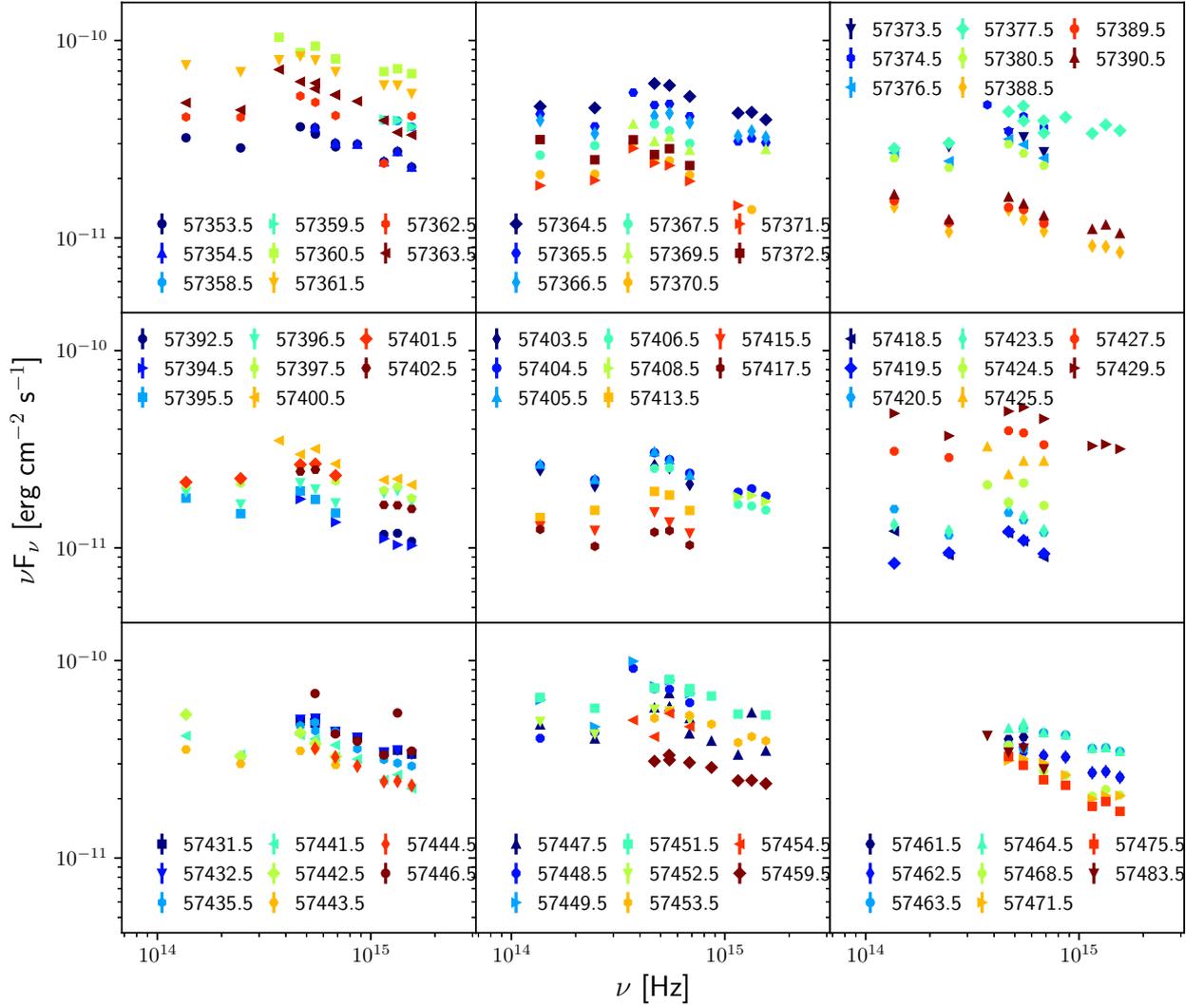}
\caption{NIR to UV SEDs of OJ 287 between MJD 57350 -- 57485 when the source
was active and the sampling was good. Only dates having observations in at least
3 different bands are shown, labeled at the mid MJD.}
\label{fig:irUVseds}
\end{figure}

In Figure \ref{fig:irUVseds}, we show 72 daily NIR to UV SEDs taken between MJD
57350 (2015 November 24) and MJD 57485 (2016 April 7). Only dates for which
we have observations in at least three different bands are plotted. Two striking
features of these SEDs are the transition from a rather steep visible spectrum to
a flatter one in the UV and the offset between the NIR and the visible portions.
The overall broad enhancement in the optical/UV can be explained as the ``big blue bump''
arising from the multi-colour emission from the accretion disc \citep{1989ApJ...346...68S,
2007A&A...473..819R} of the primary SMBH. The very high mass of the SMBH implies
rather low accretion disc temperatures, and the modest redshift further lowers the
peak wavelength of observed quasi-thermal emission (see SED modelling in \S\ref{sec:discussion}).
The additional flux in the visible
causing the break between the optical and UV portions of the SED could correspond
to a contribution from bremsstrahlung  \citep{2016ApJ...819L..37V} with some flux
arising from a ``little blue bump'' from emission lines  \citep{2007A&A...473..819R,
2017MNRAS.465.4423G}, though of course those are not strong in BL Lacs. 
The blue bump in the SED of OJ 287 does not appear to have been discussed
previously and is hard to relate to jet activity \citep[e.g.][]
{2012MNRAS.427...77V} and thus is a strong indication of a quasi-thermal, presumably,
accretion disc, component.  The modelling of the optical flare fluxes during the
first portion of this light curve that was carried out by \cite{2016ApJ...819L..37V}
indicates the presence of an underlying component that is not obviously related to
the jet and  we may be detecting its spectrum through this filter photometry. While
not uncommon in FSRQs \citep[e.g.][]{2007A&A...473..819R,2017MNRAS.469..255G}, to
our knowledge, such a feature has not been seen previously in any BL Lac object.

\begin{figure}
\centering
 \includegraphics[scale=0.83]{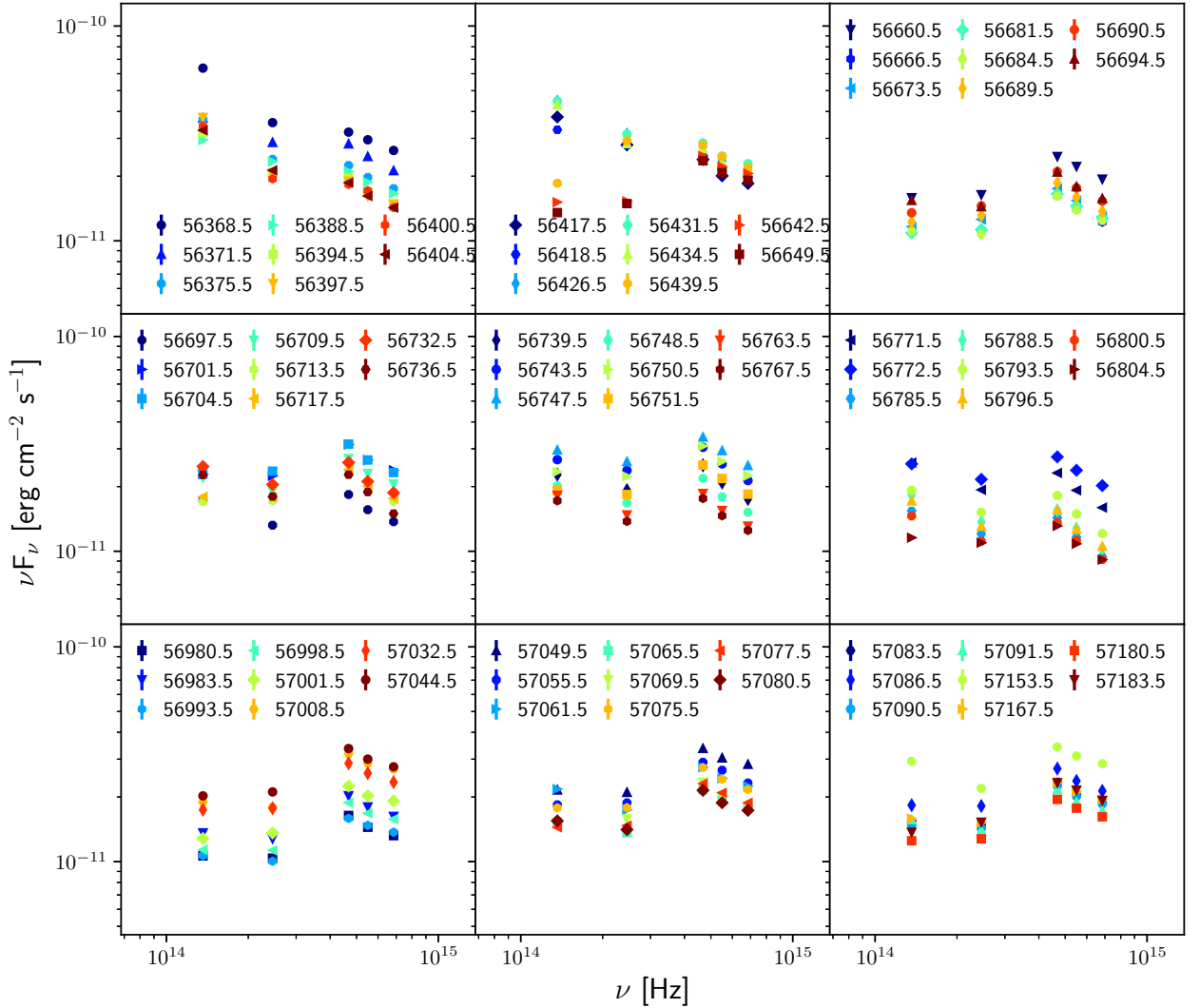}
\caption{NIR to optical SEDs of OJ 287 between MJD 56365 -- 57184, before
the intensive multi-wavelength measurements of activity around December 2015
(labels as in Fig. \ref{fig:irUVseds}).}
\label{fig:SMARTSseds}
\end{figure}

In Figure \ref{fig:SMARTSseds}, we present the NIR-optical SEDs from pre-major flare SMARTS
measurements taken between MJD 56365 (2013 Mar 14) and MJD 57184 (2015 Jun 11).
At the beginning of this period there is only a slight excess of the optical
flux above the extrapolation of the NIR fluxes, but starting on MJD 56439 (2013
May 27) there is a clear excess in the optical portion that presumably can be
attributed to a substantial accretion disc contribution that we also saw during
the November 2015 -- April 2016 data presented in Figure \ref{fig:irUVseds}.

\begin{figure}
\centering
\includegraphics[scale=1.]{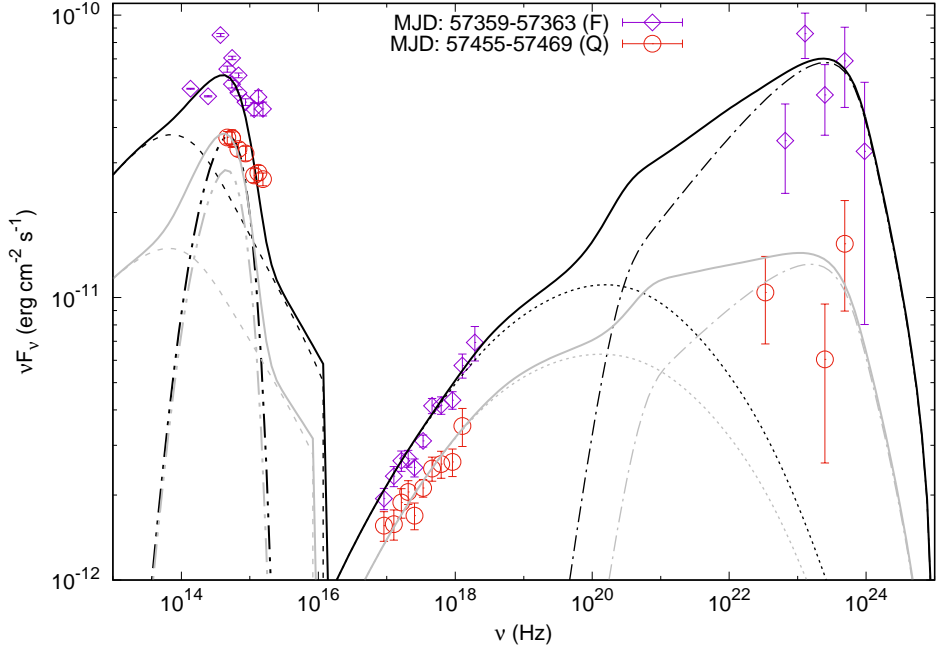}
\caption{ Broadband SEDs of  OJ 287  during two different epochs corresponding
to a flaring (F) and a quiescent state (Q) (see \S\ref{subsec:SED}). For the
model fits, the
dashed, dotted, and dotted-dashed curves represent contributions from the synchrotron, SSC and IC
of line emission (little blue bump in the UV), respectively, with black corresponding
to the F state and grey to the Q. The double-dotted-dashed curves
are the standard multi-temperature blackbody emissions from the accretion disc
associated with the primary SMBH while the solid curves are the total emission
(see \S\ref{sec:discussion}).}
\label{fig:mwSED}
\end{figure}

In Figure \ref{fig:mwSED}, we present the MW broadband SEDs (see \S\ref{sec:discussion} for
modelling) integrated over two
different epochs: the flare (F) SED (MJD: 57359--57363) data were obtained when OJ 287
had simultaneous high fluxes in all the bands (excluding radio) and the quiescent
(Q) SED (MJD: 57455--57469) that reflects a period when it was faint in the $\gamma$-rays and
rather steady in the optical--UV bands. Interestingly, the $\gamma$-ray SEDs are
very different from the previously published jet SEDs \citep[e.g.][]{2013MNRAS.433.2380K,
2009PASJ...61.1011S} and show harder spectra with a clear shift in the location of the 
peak at $\gamma$-ray energies. Also, the optical polarization associated
with the first flare is low \citep[$< 11\%$;][]{2016ApJ...819L..37V} compared to
previous flares \citep{2012ApJ...747...63A} or some of the flares following that one
(see Fig. \ref{fig:pollc}) when significant rises are seen in optical bands.

\section{Discussion}\label{sec:discussion}
We have performed a multi-wavelength correlation and spectral analysis of one of
the high activity periods of OJ 287 in December 2015, which was predicted
to occur \citep[e.g.][]{2011AcPol..51f..76V}, and is argued to be a result of
the impact of a secondary SMBH on the primary's accretion disk \citep[and references
therein; see more below]{2016ApJ...819L..37V}. The multi-band correlation
analyses were performed only for the periods when the data cadence available for 
both light curves were reasonably informative. However, the insufficiency of the
sampling is clearly visible for several cross-band correlations (see Table 1). Despite
this limitation, during temporal Segments 1 and 2 (MJD 57339--57415) when the data
cadence is quite good, the results  in Fig.\ \ref{fig:zdcf} suggest that the
variations between the NIR and  $\gamma$-ray bands are simultaneous within
the observational cadence. However, this is not the case for the $\gamma$--V/R/J
correlations during the Segment 3 (MJD 57415--57460). The delays/lags suggested
for that period by  Fig.\ \ref{fig:zdcf} are ambiguous, as while variations
are seen in both $\gamma$-ray and optical--NIR bands, the light curves do not appear
to be well correlated (Fig.\ \ref{fig:mwlc}). Further, the DCF values, even at
the nominal delayed peaks in that segment, are rather low (DCF $\lesssim$ 0.5) and
hence are formally non-significant and thus, not quoted in Table 1.
During Segment 3 there is a strong flare in the NIR--UV bands peaking at MJD 57450
that is nearly as powerful as the main flare in most of those bands, but it has no
apparent X-ray or $\gamma$-ray counterpart and also has a much higher optical polarization
than does the first flare, indicating a significant difference in its origin.
In between these two strong NIR-optical flares (peak MJDs: 57365 and 57450), there
are several other flares and almost all are associated with a significant change
in PA ($\gtrsim 90^\circ$) and PD, indicating the presence of organized magnetic
fields.

In the spectral domain, the detailed NIR--UV SEDs produced during the entire
duration of the observations when sufficient data are available across these
bands show clear  signatures  of breaks between the NIR-visible and visible-UV
spectra. The broad peak through the visible and UV part of the spectrum is good
evidence for the presence of variable thermal accretion disc emission here making a significant
contribution to the total flux although it is normally overwhelmed by the boosted
jet emission in BL Lacs.  In addition, the visible excess above the big blue bump
may be attributed to bremsstrahlung \citep[e.g. see][]{2012MNRAS.421.1861V,
2016ApJ...819L..37V}, and perhaps, some emission from a weak broad line region.
Further, the $\gamma$-ray portions of the two broad MW SEDs (see Fig. \ref{fig:mwSED}),
are relatively flat (though quite noisy) and show a shift in the location of
the peak of the SED at $\gamma$-rays, compared to the previously published SEDs
\citep[see also \citet{2012MNRAS.427...77V}]{2013MNRAS.433.2380K,2009PASJ...61.1011S}
for which the emission has been claimed to originate at parsec-scales down the jet
\citep{2013MNRAS.433.2380K,2012ApJ...747...63A}. 
 
The first optical flare, peaking on MJD 57361 corresponds to the $\sim 12$ year
near-periodicity discovered in the source over two decades ago. AGNs, in general,
are believed to be comprised of several important constituents, including SMBH(s),
accretion disc, corona, and, relativistic jets in the case of radio-loud sources,
with possibly dynamically important magnetic fields. Hence, a quasi-periodicity can,
in general, result from many processes in addition to the proposed binary SMBH model
\citep{1996ApJ...460..207L}. One such process is jet precession, and it has
been claimed in 3.5 cm radio observations of OJ 287 \citep{2004ApJ...608..149T},
but recent and better observations rather suggest a wobbling jet
\citep{2012ApJ...747...63A}. Other possible sources for
the substantial variations on a decade-like timescale include a precessing disk and
thus, jet \citep[e.g.][]{1997ApJ...478..527K}, perturbations in the inner part of
the accretion disk \citep[e.g.][]{2006ApJ...650..749L,2014MNRAS.443...58W}, coherent
helical motion of blobs in the jet \citep{1992A&A...255...59C,2015ApJ...805...91M},
and magnetic field related origins such as ``magnetic breathing'' \citep[discussion
and references therein]{2010MNRAS.402.2087V}. However, in a jet
precession related origin, one expects similar light curve profiles in all bands and
SEDs similar to those observed for the jet emission previously. Thus, although VLBA
observations suggest the possibility of a precessing jet \citep{2004ApJ...608..149T,
2012ApJ...747...63A} at parsec-scales, the appearance of a  likely thermal signature
bump does not fit this picture well. Hence, we disfavor jet precession as the
likely process for the $\sim$ 12 years quasi-periodic variability, though it could be
responsible for much longer time-scale variations as argued in \citet[see also
\citet{2010MNRAS.402.2087V}]{2012MNRAS.421.1861V}. Similarly, for a magnetic field
dominated origin, the spectra are expected to be completely non-thermal in nature.

A low PD in the case of jet/magnetic origin of the flare can result from the presence
of substantial turbulence in the emission region, but in such a case, a systematic
rotation of PA is not expected \citep[e.g.][]{2014ApJ...780...87M}. Another possibility
for producing low PD is contamination from the putative jet from the secondary SMBH,
if its polarization is opposite to that of the primary SMBH jet. This scenario can
probably lead to large PA swing as well. However, this, as well as other possible
signatures of binary nature of the source on polarization during the period of
interaction, can be explored only with time-dependent modelling. It is important
to note that the signature of the apparent thermal component is present throughout this period,
from the occurrence of the first flare on 5 December 2015 through, as long as we
had good NIR-optical-UV coverage. In fact, a detailed examination of the available
SMARTS NIR-optical data shows that the bump in the NIR-visible part appeared first 
on MJD 56439 (2013 May  27) and appears to have been present since then (Fig.\
\ref{fig:SMARTSseds}). Additionally,
this thermal signature is seen in the bands where it is expected in the case
of accretion disc around an SMBH of mass $\sim 1.8 \times 10^{10} M_{\odot}$, at the
redshift of OJ 287, thereby further supporting the SMBH binary model for OJ 287.

The apparent simultaneous variability across the entire accessible electromagnetic
spectrum (with the exception of the radio, where long lags are expected) during
the flare period and most of the rest of these observations suggest that the observed
quasi-thermal and non-thermal emission are either co-spatial or within a light-day or
so apart from the disk. As OJ 287 is a blazar, the non-thermal emission can be associated
with the jet of the primary, or even the secondary, SMBH. The latter has been argued to be
probably aligned with the spin axis of the primary SMBH, and hence, should also have its putative
jet pointed close toward us \citep{2016ApJ...819L..37V}. On the other hand, the inferred
simultaneous variability and the hard $\gamma$-ray spectra compared to those measured
previously \citep[e.g.][]{2013MNRAS.433.2380K}, along with a shift in the SED peak, favors a
leptonic origin of the second SED bump emission,  with $\gamma$-rays resulting from
IC of emission line photons, i.e., the little blue bump (optical-UV) that was never
seen in previous studies.

As shown in Fig. \ref{fig:mwSED}, modelling of SEDs with a one-zone model, assuming
synchrotron and IC scattering of both synchrotron and little blue bump photons from
a broken power-law particle distribution, can successfully reproduce the observed
broadband SEDs at the two epochs \citep[e.g. see][for more details on the modelling
approach]{2013MNRAS.433.2380K}. In such one-zone models, the optically thin
synchrotron (NIR-optical) and SSC (X-ray spectrum, away from the peak of the SED)
emissions reflect the instantaneous particle spectrum. Thus the particle spectrum before
the break is derived from the X-ray data, but due to thermal contamination at the
NIR-optical part, we have used the quiescent state particle index 
from a previous study of the OJ 287 by \citet{2013MNRAS.433.2380K}, whose approach
we follow here.
 For an assumed particle spectrum, emission region size, and  jet
angle to the line of sight ($\theta$), the flux due to IC of external field before
the peak of the SED is (in a delta function approximation for the IC and synchrotron
emission)
\begin{equation}\label{eq:ec}
 F(\nu_{\gamma}) \propto [\Gamma(1+cos\theta)]^{(p+1)/2} [\nu_\star \delta]^{(p+5)/2} K \nu_\gamma^{-(p-1)/2},
\end{equation}
where $\rm \Gamma,~\delta~(=[\Gamma(1-cos\theta)]^{-1}),~K$, and $\nu_\star$ are respectively,
the jet bulk Lorentz factor, Dopper factor, particle normalization, and the photon
field for the IC scattering (little blue bump) with $p$ and $q$ (see below) being the indices before
and after the break ($\rm \gamma_b$) in the particle spectrum. Similarly, the synchrotron flux
after the SED peak in the optical-NIR region is 
\begin{equation}\label{eq:syn}
 F(\nu) \propto \delta^{(q+5)/2} B^{(p+1)/2} K \gamma_b^{(q-p)} \nu^{-(q-1)/2},
\end{equation}
where $B$ is the magnetic field in the emission region and the SED peak of the synchrotron
emission is related to Larmor frequency ($\rm \nu_L$) by
\begin{equation}\label{eq:synPeak}
 \nu_p^{syn} \approx \frac{\delta}{(1+z)} \gamma_b^2 \nu_L.
\end{equation}
Thus, for a given $\rm \delta$, $K$ can be estimated using Eqn.\ \ref{eq:ec} for a
given observed flux at a $\gamma$-ray energy while constraining $\gamma_b$ by demanding the
location of the SED peak below NIR K-band using Eqn.\ \ref{eq:synPeak}. With this,
the magnetic field can be derived using Eqn.\ \ref{eq:syn} such that the sum of
synchrotron and thermal contribution reproduces the observed break at the NIR-optical
part and the X-ray spectrum by SSC. We choose a $\delta$ value so that the system
remains in equipartition\footnote{particle energy density equals magnetic energy density}.
The corresponding model parameters for the assumed particle spectrum are given in Table \ref{tab:parSED} where
the size of the emission region has been kept the same for both flare and quiescent
epochs. Due to the onset of the Klein-Nishina effect for the IC emission, the $\gamma$-ray spectra
is essentially independent of the assumed NIR-optical particle spectrum. For the NIR-optical part of the SED,
the non-thermal and thermal contributions are constrained so that the sum of those contributions
does not exceed the observed data during the ``F'' state at NIR bands (K).  However, for
the ``Q'' state, due to lack of any data in NIR-bands then, we have simply scaled down
the synchrotron part as well as the thermal part of the emission. The accretion-disk portion of
the SED, on the other hand, is reproduced employing multi-temperature blackbody
emission from the primary SMBH ($\rm M \simeq 1.8\times10^{10}~M_\odot$) assuming a
10\% accretion efficiency. Our modelling suggests that the accretion rate is modestly
temporally variable; however, one needs better constraints on the non-thermal 
contributions to more precisely estimate the accretion rates. Additionally, the
simultaneous MW variability along with $\gamma$-rays arising from IC scattering
of the little blue bump photons also constrain the location of the emission region
to sub-parsec scales, i.e., within the line emission region, contrary to previous
inferences of origin at parsec scales \citep{2017A&A...597A..80H,2013MNRAS.433.2380K,
2012ApJ...747...63A}. Furthermore, for the given parameters, the IC of a putative 250 K
blackbody field \citep{2013MNRAS.433.2380K}  makes a negligible contribution.  
Bremsstrahlung, and/or an additional synchrotron component, being broadband emission,
would not lead to the structure seen in
the NIR, where the I-band excess is most likely the H$\alpha$ line-emission at the redshift
of OJ 287, (e.g. \citet{2007A&A...473..819R} for the J-band excess in 3C 454.3). 

\begin{table}
\centering
\caption{ SED parameters}
\begin{tabular}{l c c}
\hline
Parameters & F & Q  \\
\hline
Particle index before break (p)  & 2.36 & 2.50 \\
(from X-ray) &  &  \\
Particle index after break (q) & 3.8 & 3.8 \\
Magnetic field (Gauss) & 0.9 & 1.1 \\
particle break energy $(\gamma_b^\ast)$ & 1590 & 1721 \\
Equipartition fraction$^{\ast\ast}$ & 1 & 1 \\ 
Doppler factor & 14 & 10 \\
Jet power (erg/s) & $\rm 3.0 \times 10^{45}$ & $\rm 2.3 \times 10^{45}$ \\
\hline
\multicolumn{3}{l}{Size of the emission region: $\rm 3 \times 10^{16}$ cm} \\
\multicolumn{3}{l}{Jet angle to the line of sight: $\rm 3^\circ$} \\
\multicolumn{3}{l}{Minimum and maximum electron Lorentz factor: 40, $3\times 10^{4}$} \\
\multicolumn{3}{l}{$^\ast$in units of electron rest mass energy} \\
\multicolumn{3}{l}{$^{\ast\ast}$ particle-energy-density/magnetic-energy-density} \\
\end{tabular}
\label{tab:parSED}
\end{table}

The near periodic occurrence of optical outbursts from OJ 287 every $\sim$12 years
and the associated multi-wavelength variability makes it an outstanding
candidate for increasing our  understanding of various aspects of disc-jet connections,
other jet processes and various proposed models for them \citep[e.g.][and discussion
therein]{2010MNRAS.402.2087V}, even if it is an apparently unique case. Since the
times of the flares  can be predicted, observations can be scheduled for maximum
possible coverage across the electromagnetic spectrum, which is much more difficult
to do for normal blazars with their erratic fluctuations. The multi-wavelength
variability data, including optical polarization information, presented here, along
with that taken elsewhere for part of this flare \citep[see][]{2016ApJ...819L..37V,
2017MNRAS.465.4423G}, as well as for past and future outbursts,  promise to answer
some of the current issues related to understanding jets and accretion discs.

\section{Conclusions}\label{sec:conclude}
We performed a multi-wavelength spectral and temporal correlation analyses
of emission from OJ 287 associated with its recent bright optical flaring activity observed
in December 2015. The MW light curves from December 2015 -- May 2016 show
significant activity in all bands, most prominently in NIR, optical, UV,
and X-ray, with most of them associated with a significant change in PA
and PD. The temporal variations are simultaneous in all these bands to within the
observational cadences whenever there are sufficient data to properly estimate the lags.

In the spectral domain, several new features were observed
in the NIR-UV and gamma-ray bands that have not been seen in previous studies
of this source nor even in any other BL Lac source. Most importantly, the NIR-optical-UV
SEDs show bumps in the NIR-optical and optical-UV with hints of excess in the I-band.
The NIR-optical bump is consistent with a multi-temperature accretion disc
emission from the primary SMBH of mass $\sim 1.8\times10^{10} M_\odot$, while
the optical-UV bump seems consistent with line emission, probably a result of
the observed heightened disc emission. Interestingly, the broadband SEDs extracted during a high activity
state and a low activity state are also quite different from those of previously
studied LAT-band SEDs at  low and high states of OJ 287. Compared
to previous SEDs, the present SEDs  show hardening and a shift in the
SED peak at $\gamma$-rays. The shift and a harder $\gamma$-ray spectrum are consistent with IC
scattering of photons from the little-blue-bump, i.e., line-emission. This, along with
 the lack of any apparent lags between bands, suggest that the dominant emission
region, at least during this period,  is
located at sub-parsec scales and so within the broad line region. The likely thermal
bump in the visible/UV, which can be traced back to MJD 56439 using available NIR-optical
data, coupled with the simultaneous multi-wavelength variability and relatively weak
$\gamma$-ray emission seen during the period MJD 57310--57560 that we have focused upon here,
favor the already well supported binary SMBH model. 

\section*{Acknowledgements}
The authors thank the referee Manasvita Joshi for valuable comments and suggestions.
This research has made use of data, software and web tools of High Energy Astrophysics 
Science Archive Research Center (HEASARC), maintained by NASA's Goddard Space Flight 
Center and up-to-date SMARTS optical/near-infrared light curves available at \\
www.astro.yale.edu/smarts/glast/home.php. SMARTS observations of Large Area 
Telescope monitored blazars are supported by Yale University and Fermi GI grant 
NNX 12AP15G, and the SMARTS 1.3-m observing queue received support from NSF grant 
AST 0707627. Data from the Steward Observatory spectropolarimetric monitoring project 
were used. This programme is supported by Fermi Guest Investigator grants NNX08AW56G, 
NNX09AU10G, NNX12AO93G, and NNX15AU81G. The OVRO 40 m Telescope Fermi Blazar Monitoring 
Program is supported by NASA under awards NNX08AW31G and NNX11A043G, and by the NSF under 
awards AST-0808050 and AST-1109911. 

PK acknowledges support from FAPESP grant no.\ 2015/13933-0. ACG's work is partially
supported by Chinese Academy of Sciences (CAS) President's International Fellowship
Initiative (PIFI) grant no.\ 2016VMB073.  PJW is grateful for hospitality at KIPAC,
Stanford University, and SHAO during a sabbatical. HG is sponsored by a CAS Visiting
Fellowship for Researchers from  Developing Countries, CAS PIFI (grant no.\ 2014FFJB0005),
supported by the NSFC Research Fund for International Young Scientists (grant no.
\ 11450110398) and supported by a Special Financial Grant from the China Postdoctoral
Science Foundation (grant no.\ 2016T90393). EMGDP acknowledges support from
the Brazilian funding agencies FAPESP (grant 2013/10559-5) and CNPq (grant 306598/2009-4).
The Abastumani team acknowledges financial
support by the Shota Rustaveli National Science Foundation under contract FR/217950/16.
The St.\ Petersburg University team acknowledges support from Russian RFBR grant 15-02-00949 
and St.\ Petersburg University research grant 6.38.335.2015. GD and OV gratefully acknowledge 
the observing grant support from the Institute of Astronomy and Rozhen National Astronomical 
Observatory, Bulgaria Academy of Sciences, via bilateral joint research project ``Observations 
of ICRF radio-sources visible in optical domain" (the head is GD). This work is a part of the 
Projects No. 176011 (Dynamics and kinematics of celestial bodies and systems), No 176004
(Stellar physics) and No 176021 (Visible and invisible matter in nearby galaxies: theory and 
observations) supported by the Ministry of Education, Science and Technological Development 
of the Republic of Serbia.
The work of ES, AS, RB was partially supported by Scientific Research 
Fund of the Bulgarian Ministry of Education and Sciences under grant DN 08-1/2016. MFG is 
supported by the National Science Foundation of China (grants 11473054 and U1531245) and by the 
Science and Technology Commission of Shanghai Municipality (grant 14ZR1447100). ZZ is thankful 
for support from the CAS Hundred-Talented program (Y787081009). JHF's work is supported by the 
National Natural Science Foundation of China (NSFC U1531245) and  Guangdong Innovation Team for 
Astrophysics (2014KCXTD014).


\begin{thebibliography}{}

\bibitem[\protect\citeauthoryear{Acero et al.}{2015}]{2015ApJS..218...23A} Acero F.,
et al., 2015, ApJS, 218, 23

\bibitem[\protect\citeauthoryear{Agarwal et al.}{2015}]{2015MNRAS.451.3882A}
Agarwal A., et al. 2015, MNRAS, 451, 3882

\bibitem[\protect\citeauthoryear{Agudo et al.}{2011}]{2011ApJ...726L..13A} Agudo I.,
et al., 2011, ApJ, 726, L13 

\bibitem[\protect\citeauthoryear{Agudo et al.}{2012}]{2012ApJ...747...63A} Agudo I.,
Marscher A.~P., Jorstad S.~G., G{\'o}mez J.~L., Perucho M., Piner B.~G., Rioja M.,
Dodson R., 2012, ApJ, 747, 63 

\bibitem[\protect\citeauthoryear{Alexander}{2013}]{2013arXiv1302.1508A} 
Alexander T., 2013,  arXiv:1302.1508 

\bibitem[\protect\citeauthoryear{Alexander}{1997}]{1997ASSL..218..163A} 
Alexander T., 1997, ASSL, 218, 163

\bibitem[\protect\citeauthoryear{Atwood et al.}{2009}]{2009ApJ...697.1071A} 
Atwood W.~B., et al., 2009, ApJ, 697, 1071

\bibitem[\protect\citeauthoryear{Bessell, Castelli, \& Plez}{1998}]
{1998A&A...333..231B} Bessell M.~S., Castelli F., Plez B., 1998, A\&A, 333, 231

\bibitem[\protect\citeauthoryear{Bhatta et al.}{2016}]{2016ApJ...832...47B}
Bhatta G., et al., 2016, ApJ, 832, 47

\bibitem[\protect\citeauthoryear{Camenzind \& Krockenberger}{1992}]{1992A&A...255...59C}
Camenzind M., Krockenberger M., 1992, A\&A, 255, 59

\bibitem[\protect\citeauthoryear{Fossati et al.}{1998}]{1998MNRAS.299..433F} Fossati
G., Maraschi L., Celotti A., Comastri A., Ghisellini G., 1998, MNRAS, 299, 433

\bibitem[\protect\citeauthoryear{Ghisellini et al.}{2017}]{2017MNRAS.469..255G}
Ghisellini G., Righi C., Costamante L., Tavecchio F., 2017, MNRAS, 469, 255
	
\bibitem[\protect\citeauthoryear{Gupta et al.}{2017}]{2017MNRAS.465.4423G} Gupta
A.~C., et al., 2017, MNRAS, 465, 4423

\bibitem[\protect\citeauthoryear{Healey et al.}{2008}]{2008ApJS..175...97H}
Healey S.~E. et al., 2008, ApJS, 175, 97

\bibitem[\protect\citeauthoryear{Hodgson et al.}{2017}]{2017A&A...597A..80H} Hodgson
J.~A., et al., 2017, A\&A, 597, A80

\bibitem[\protect\citeauthoryear{Hudec et al.}{2013}]{2013A&A...559A..20H}
Hudec R., Ba{\v s}ta M., Pihajoki P., Valtonen M., 2013, A\&A, 559, A20


\bibitem[\protect\citeauthoryear{Katz}{1997}]{1997ApJ...478..527K} Katz J.~I.,
1997, ApJ, 478, 527

\bibitem[\protect\citeauthoryear{Kushwaha, Sahayanathan, \& Singh}{2013}]{2013MNRAS.433.2380K}
Kushwaha P., Sahayanathan S., Singh K.~P., 2013, MNRAS, 433, 2380 

\bibitem[\protect\citeauthoryear{Kushwaha, Singh, \& Sahayanathan}{2014}]{2014ApJ...796...61K}
Kushwaha P., Singh K.~P., Sahayanathan S., 2014, ApJ, 796, 61 

\bibitem[\protect\citeauthoryear{Larionov et al.}{2008}]{2008A&A...492..389L}
Larionov V.~M., et al. 2008, A\&A, 492, 389

\bibitem[\protect\citeauthoryear{Lehto \& Valtonen}{1996}]{1996ApJ...460..207L}
Lehto H.~J., Valtonen M.~J., 1996, ApJ, 460, 207

\bibitem[\protect\citeauthoryear{Lister et al.}{2013}]{2013AJ....146..120L} Lister
M.~L., et al., 2013, AJ, 146, 120 

\bibitem[\protect\citeauthoryear{Liu, Zhao, \& Wu}{2006}]{2006ApJ...650..749L}
Liu F.~K., Zhao G., Wu X.-B., 2006, ApJ, 650, 749 

\bibitem[\protect\citeauthoryear{Mannheim \& Biermann}{1992}]{1992A&A...253L..21M} Mannheim K.,
Biermann P.~L., 1992, A\&A, 253, L21

\bibitem[\protect\citeauthoryear{Mao et al.}{2016}]{2016ApJS..224...26M} Mao P.,
Urry C.~M., Massaro F., Paggi A., Cauteruccio J., K{\"u}nzel S.~R., 2016, ApJS, 224, 26

\bibitem[\protect\citeauthoryear{Marscher}{2014}]{2014ApJ...780...87M} Marscher
A.~P., 2014, ApJ, 780, 87

\bibitem[\protect\citeauthoryear{Meyer et al.}{2017}]{2017arXiv170105846M} Meyer
E.~T., et al., 2017, arXiv:1701.05846

\bibitem[\protect\citeauthoryear{Mohan \& Mangalam}{2015}]{2015ApJ...805...91M}
Mohan P., Mangalam A., 2015, ApJ, 805, 91

\bibitem[\protect\citeauthoryear{Mo{\'o}r et al.}{2011}]{2011AJ....141..178M}
Mo{\'o}r A., Frey S., Lambert S.~B., Titov O.~A., Bakos J., 2011, AJ, 141, 178


\bibitem[\protect\citeauthoryear{Moretti et al.}{2005}]{2005SPIE.5898..360M}
Moretti A., et al., 2005, SPIE, 5898, 360

\bibitem[\protect\citeauthoryear{M{\"u}cke \& Protheroe}{2001}]{2001APh....15..121M}
M{\"u}cke A., Protheroe R.~J., 2001, APh, 15, 121

\bibitem[\protect\citeauthoryear{Pihajoki}{2016}]{2016MNRAS.457.1145P} Pihajoki P.,
2016, MNRAS, 457, 1145

\bibitem[\protect\citeauthoryear{Pihajoki et al.}{2013}]{2013ApJ...764....5P}
Pihajoki P., et al., 2013, ApJ, 764, 5

\bibitem[\protect\citeauthoryear{Pihajoki, Valtonen, \& Ciprini}{2013}]
{2013MNRAS.434.3122P} Pihajoki P., Valtonen M., Ciprini S., 2013, MNRAS, 434, 3122

\bibitem[\protect\citeauthoryear{Potter \& Cotter}{2013}]{2013MNRAS.429.1189P}
Potter W.~J., Cotter G., 2013, MNRAS, 429, 1189 

\bibitem[\protect\citeauthoryear{Raiteri et al.}{2007}]{2007A&A...473..819R}
Raiteri C.~M. et al., 2007, A\&A, 473, 819

\bibitem[\protect\citeauthoryear{Rakshit et al.}{2017}]{2017ApJ...835..275R} Rakshit S.,
Stalin C.~S., Muneer S., Neha S., Paliya V.~S., 2017, ApJ, 835, 275

\bibitem[\protect\citeauthoryear{Roming et al.}{2009}]{2009ApJ...690..163R} Roming
P.~W.~A., et al., 2009, ApJ, 690, 163

\bibitem[\protect\citeauthoryear{Roming et al.}{2005}]{2005SSRv..120...95R} Roming
P.~W.~A., et al., 2005, SSRv, 120, 95

\bibitem[\protect\citeauthoryear{Sandrinelli et al.}{2016}]{2016AJ....151...54S}
Sandrinelli A., Covino S., Dotti M., Treves A., 2016, AJ, 151, 54

\bibitem[\protect\citeauthoryear{Sawada-Satoh et al.}{2015}]{2015PKAS...30..429S}
Sawada-Satoh S., et al., 2015, PKAS, 30, 429

\bibitem[\protect\citeauthoryear{Schlafly \& Finkbeiner}{2011}]{2011ApJ...737..103S}
Schlafly E.~F., Finkbeiner D.~P., 2011, ApJ, 737, 103

\bibitem[\protect\citeauthoryear{Schlegel, Finkbeiner, \& Davis}{1998}]{1998ApJ...500..525S}
Schlegel D.~J., Finkbeiner D.~P., Davis M., 1998, ApJ, 500, 525

\bibitem[\protect\citeauthoryear{Seta et al.}{2009}]{2009PASJ...61.1011S} Seta H.,
et al., 2009, PASJ, 61, 1011

\bibitem[\protect\citeauthoryear{Sillanp{\"a}{\"a} et al.}{1996a}]{1996A&A...305L..17S}
Sillanp{\"a}{\"a}  A., et al., 1996a, A\&A, 305, L17

\bibitem[\protect\citeauthoryear{Sillanp{\"a}{\"a}  et al.}{1996b}]{1996A&A...315L..13S}
Sillanp{\"a}{\"a}  A., et al., 1996b, A\&A, 315, L13

\bibitem[\protect\citeauthoryear{Sillanp{\"a}{\"a}  et al.}{1988}]{1988ApJ...325..628S}
Sillanp{\"a}{\"a}  A., Haarala S., Valtonen M.~J., Sundelius B., Byrd G.~G., 1988, ApJ, 325, 628

\bibitem[\protect\citeauthoryear{Smith et al.}{2009}]{2009arXiv0912.3621S} 
Smith P.~S., Montiel E., Rightley S., Turner J., Schmidt G.~D., Jannuzi 
B.~T., 2009,  2009 Fermi Symposium, eConf Proc.\ C091122: arXiv:0912.3621

\bibitem[\protect\citeauthoryear{Sun \& Malkan}{1989}]{1989ApJ...346...68S}
Sun W.-H. \& Malkan M.~A. 1989, ApJ, 346, 68

\bibitem[\protect\citeauthoryear{Tateyama}{2013}]{2013ApJS..205...15T} 
Tateyama C.~E., 2013, ApJS, 205, 15

\bibitem[\protect\citeauthoryear{Tateyama \& Kingham}{2004}]{2004ApJ...608..149T}
Tateyama C.~E., Kingham K.~A., 2004, ApJ, 608, 149


\bibitem[\protect\citeauthoryear{Valtonen \& Sillanp{\"a}{\"a}}{2011}]
{2011AcPol..51f..76V} Valtonen M., Sillanp{\"a}{\"a} A., 2011, AcPol, 51, 060000 

\bibitem[\protect\citeauthoryear{Valtonen \& Wiik}{2012}]{2012MNRAS.421.1861V}
Valtonen M.~J., Wiik K., 2012, MNRAS, 421, 1861

\bibitem[\protect\citeauthoryear{Valtonen, Ciprini, \& Lehto}{2012}]{2012MNRAS.427...77V}
Valtonen M.~J., Ciprini S., Lehto H.~J., 2012, MNRAS, 427, 77

\bibitem[\protect\citeauthoryear{Valtonen et al.}{2006}]{2006ApJ...646...36V}
Valtonen M.~J., et al., 2006, ApJ, 646, 36

\bibitem[\protect\citeauthoryear{Valtonen et al.}{2011}]{2011ApJ...742...22V} Valtonen M.~J.,
Mikkola S., Lehto H.~J., Gopakumar A., Hudec R., Polednikova J., 2011, ApJ, 742, 22

\bibitem[\protect\citeauthoryear{Valtonen et al.}{2016}]{2016ApJ...819L..37V}
Valtonen M.~J., et al., 2016, ApJ, 819, L37

\bibitem[\protect\citeauthoryear{Villforth et al.}{2010}]{2010MNRAS.402.2087V}
Villforth C., et al., 2010, MNRAS, 402, 2087

\bibitem[\protect\citeauthoryear{Wang et al.}{2014}]{2014MNRAS.443...58W} Wang J.-Y.,
An T., Baan W.~A., Lu X.-L., 2014, MNRAS, 443, 58 

\bibitem[\protect\citeauthoryear{Wu et al.}{2006}]{2006AJ....132.1256W} Wu J.,
et al., 2006, AJ, 132, 1256


\end{thebibliography}
\end{document}